\documentstyle[preprint,prd,aps]{revtex}
\begin{document}
\draft
%\twocolumn[\hsize\textwidth\columnwidth\hsize\csname @twocolumnfalse\endcsname

\title{Constraining Supersymmetric SO(10) Models Through Cosmology}
\author{Rachel Jeannerot and Anne-Christine Davis\thanks{and King's College,
Cambridge} ,\\
        {\normalsize{ Department of Applied Mathematics and Theoretical
Physics, Cambridge University,}}\\
        {\normalsize{ Silver Street, Cambridge, CB3 9EW, UK}}}
\date{\today}
\maketitle
\begin{abstract}
We study the impact of the symmetry breaking patterns from
supersymmetric SO(10) down to the standard model on the standard
big-bang cosmology through the formation of topological defects. None of
the models is consistent with the standard cosmology without invoking
any mechanism to solve the monopole problem. For this purpose, we use a
hybrid false vacuum inflationary scenario. Only two symmetry
breaking patterns are consistent with these topological considerations
and with the actual data on the proton lifetime.
\end{abstract}
%\vskip2pc]

\section{introduction}

Grand Unified Theories (GUTs) have been constructed to unify the strong,
 weak and electromagnetic interactions. The minimal grand unified group
in which all
kinds of matter are unified is SO(10) GUT \cite{Georgi}. Indeed SO(10)
has a 16 dimensional spinorial representation and therefore all quarks
and leptons belonging to a single family can be assigned to a single
multiplet. Now, when looking at the measured values at LEP of the
three gauge coupling constants and interpolating them to high
energies, we find that they do not merge. On the other hand, the three
coupling constants in the Minimal Supersymmetric Standard Model, with
supersymmetry broken at $T \sim 10^3$ GeV, merge in a single point at
$T \sim 10^{16}$ GeV \cite{merge}. Supersymmetry can also solve the
gauge hierarchy problem.

Supersymmetric SO(10) is consistent with the measured values of
$\sin^2 \theta_w$ and $\alpha_s$ and the unification of the three
gauge coupling constants at $\sim 10^{16}$ GeV \cite{merge}. It also
beautifully solves the question of fermions masses \cite{dimo}.
Furthermore it leads to a relation for $\tan \beta$, an unknown factor
within the Minimal Supersymmetric Standard Model, giving $\tan \beta =
m_t/m_b$ \cite{tanB}. Natural doublet-triplet splitting can be
achieved in supersymmetric SO(10) via the Dimopoulos-Wilczek mechanism
\cite{DW}. SO(10) also contains an unbroken matter parity which lies
in the centre of SO(10). The latter can suppress rapid proton decay
and provide a good cold dark matter candidate in the form of the
lightest superparticle. Now, introducing a 126 and a $\overline{126}$
into the supersymmetric model, the see-saw mechanism can be implemented
\cite{seesaw}, thus providing a good hot dark matter candidate;
the right-handed neutrino gets a superheavy Majorana mass and the
left-handed neutrino gets a very small mass. Supersymmetric SO(10) can
also explain the solar neutrino problem via the MSW mechanism
\cite{MSW}. Finally, it is a good candidate for baryogenesis
\cite{baryon}.

Thus, supersymmetric SO(10) is very attractive from a particle physics
point of view and can also help to solve some cosmological problems.
One would therefore like to be able to select one of the
breaking patterns. Unfortunately, there is considerable freedom in
doing so, and the only way out from a particle physics point of view
would be from string compactification.

However, any particle physics model is irrelevant if it does not
satisfy cosmological considerations. Conversely, any cosmological model is
irrelevant if it does not agree with particle physics considerations.
In other words, any GUT model is tied up with cosmology
, and one should not be considered without the other; as
nice as a GUT (respectively cosmological) model can be, it can however
lead to a cosmological catastrophe (cannot be implemented in any viable
particle physics model), and should therefore be regarded with suspicion
. When symmetries spontaneously break down, according to
Kibble mechanism \cite{Kibble}, topological defects form, such as monopoles,
strings
or domain walls. Monopoles, because they would be too abundant, and domain
walls, because they are too heavy, if present today would dominate the
energy density of the universe and lead to a cosmological catastrophe.
On the other hand, cosmic strings can explain structure formation and
part of the baryon asymmetry of the universe.

We derive below the cosmological constraints on the symmetry breaking
schemes of
supersymmetric SO(10) down to the standard model due to the formation
of topological defects. In sec.\ref{sec-breakings} we list the possible
symmetry breaking pattern involving at most one
intermediate symmetry breaking scale. In sec. \ref{sec-topo}, we review the
conditions for the formation of topological defects,
giving systematic conditions in
supersymmetric SO(10). In sec. \ref{sec-inflation} we discuss the
hybrid inflationary scenario which can be implemented in supersymmetric
SO(10). In sections \ref{sec-su5}, \ref{sec-three} and
\ref{sec-direct} we give a systematic analysis of the cosmological
implications for the different symmetry breaking scenarios listed
in section \ref{sec-breakings}. We conclude
in section \ref{sec-concl}, pointing out the only models not in conflict
with the standard cosmology.

\section{Breaking down to the standard model}
\label{sec-breakings}
In this section, we give a list of all the symmetry breaking patterns from
supersymmetric SO(10) down to the standard model, using no more than one
intermediate breaking scale. The main differences between supersymmetric
 and non-supersymmetric SO(10) models is in the symmetry breaking scales as
 we shall see and in the choice for the intermediate symmetry groups.
In non-supersymmetric models, at least one intermediate symmetry
breaking is needed in order to obtain consistency with the measured value
of $\sin^2 \theta_w$ and with the gauge coupling constants interpolated
to high energy to meet around $10^{15}$ GeV. On the other hand, in
supersymmetric SO(10) models, we can break directly down to the
standard model, breaking supersymmetry at $\sim 10^3 \: GeV$,
predicting the measured value of $\sin^2 \theta w$ and having
the gauge coupling constant joining in a single point at $2
\times 10^{16} \: GeV$.

We shall consider the following symmetry breaking patterns from
supersymmetric SO(10) down to the standard model:
\begin{eqnarray}
{\bf 1.} \qquad &SO(10) &\stackrel{M_{GUT}}{\rightarrow} SU(5) \times U(1)_X
\stackrel{M_G}{\rightarrow}  SM \nonumber \\
{\bf 2.} \qquad &SO(10) &\stackrel{M_{GUT}}{\rightarrow} SU(5)
\stackrel{M_G}{\rightarrow}  SM \nonumber \\
{\bf 3.} \qquad  & SO(10) & \stackrel{M_{GUT}}{\rightarrow} SU(5) \times
\widetilde{U(1)} \stackrel{M_G}{\rightarrow}  SM \nonumber \\
{\bf 4.}\qquad &SO(10)& \stackrel{M_{GUT}}{\rightarrow} SU(4)_c \times
SU(2)_L \times
SU(2)_R  \stackrel{M_G}{\rightarrow} SM \nonumber \\
{\bf 5.} \qquad &SO(10) &\stackrel{M_{GUT}}{\rightarrow}SU(3)_c \times
SU(2)_L \times
SU(2)_R \times U(1)_{B\! -\! L} \nonumber  \stackrel{M_G}{\rightarrow}
SM \nonumber \\
{\bf 6.}\qquad &SO(10) &\stackrel{M_{GUT}}{\rightarrow} SU(3)_c \times SU(2)_L
\times
U(1)_R \times U(1)_{B\! -\! L}   \stackrel{M_G}{\rightarrow} SM
\nonumber\\
{\bf 7.}\qquad &SO(10)& \stackrel{M_{GUT}}{\rightarrow} SM \nonumber\\
{\bf 8.} \qquad &SO(10) &\stackrel{M_{GUT}}{\rightarrow} SU(5) \times U(1)_X
\stackrel{M_G}{\rightarrow}  SM \times Z_2\nonumber \\
{\bf 9.}\qquad  &SO(10) &\stackrel{M_{GUT}}{\rightarrow} SU(5) \times
 Z_2 \\
{\bf 10.} \qquad  & SO(10) & \stackrel{M_{GUT}}{\rightarrow} SU(5) \times
\widetilde{U(1)}
\stackrel{M_G}{\rightarrow} SM \times Z_2 \nonumber\\
{\bf 11.}\qquad &SO(10) &\stackrel{M_{GUT}}{\rightarrow} SU(4)_c \times SU(2)_L
\times
SU(2)_R  \stackrel{M_G}{\rightarrow} SM
\times Z_2 \nonumber \\
{\bf 12.}\qquad &SO(10) &\stackrel{M_{GUT}}{\rightarrow}SU(3)_c \times SU(2)_L
\times
SU(2)_R \times U(1)_{B\! -\! L} \stackrel{M_G}{\rightarrow} SM
\times Z_2\nonumber\\
{\bf 13.}\qquad &SO(10) &\stackrel{M_{GUT}}{\rightarrow} SU(3)_c \times
SU(2)_L \times
U(1)_R \times U(1)_{B\! -\! L}  \stackrel{M_G}{\rightarrow} SM \times
Z_2\nonumber\\
{\bf 14.}\qquad &SO(10) &\stackrel{M_{GUT}}{\rightarrow}  SM \times Z_2
\nonumber\\
\end{eqnarray}
where $SM$ stands for the standard model gauge group $SU(3)_c \times
 SU(2)_L \times U(1)_Y$. In models ${\bf 1.}$
to ${\bf 6.}$, we break SUSY at $\sim 10^3$ GeV, and the symmetry
group
$SU(3)_c \times SU(2)_L \times U(1)_Y$ down to $SU(3)_c \times U(1)_Q$
at $\sim M_Z$. In models ${\bf 7.}$ to ${\bf 11.}$, we also break SUSY at $\sim
10^3$ GeV, and we break the group symmetry $SU(3)_c \times SU(2)_L
\times U(1)_Y \times Z_2$ down to $SU(3)_c \times U(1)_Q \times Z_2$
at $\sim M_Z$.  In the latter cases, the $Z_2$ symmetry remains
unbroken down to low energy, and acts as matter parity. It preserves
large values for the proton lifetime and stabilizes the Lightest
SuperParticle (LSP), thus
providing a good hot dark matter candidate.

In order to satisfy LEP data, we must have $M_{GUT} \sim M_G$ (see Langacker
and Luo in Ref.\cite{merge}). For non supersymmetric models, the value of
the $B \! - \! L$ symmetry breaking scale is anywhere between $10^{10}$ to
$10^{13.5}$ GeV \cite{moha}. For the supersymmetric case it is around
$10^{15}$ to $10^{16}$ GeV. Indeed, the scale $M_G$ is fixed by the
unification of the gauge couplings, and in the absence of particle
threshold corrections is $M_G \sim 10^{16}$ GeV \cite{merge}. But, as
in the non-supersymmetric case, threshold corrections can induce
uncertainties of a factor $10^{\pm 1}$ GeV. These corrections
vary with the
intermediate subgroup considered, but in any cases, we can assume that
$M_G \sim 10^{15} - 10^{16}$ GeV. The scale $M_{GUT}$ must be greater than
the unified scale $M_G$ and below the Planck scale, therefore we must
have $10^{19} GeV \geq M_{GUT} \geq 10^{15} - 10^{16} GeV$.

In order to simplify the notation, we shall use the following
\begin{eqnarray}a. \qquad && 4_c 2_L 2_R \equiv SU(4)_c \times SU(2)_L \times
SU(2)_R  \nonumber \\
b. \qquad &&  3_c 2_L 2_R 1_{B-L} \equiv SU(3)_c \times SU(2)_L \times SU(2)_R
\times U(1)_{B-L}   \nonumber \\
c. \qquad &&  3_c 2_L 1_R 1_{B-L} \equiv SU(3)_c \times SU(2)_L \times U(1)_R
\times U(1)_{B-L} \nonumber \\
d. \qquad &&   3_c 2_L 1_Y (Z_2) \equiv SU(3)_c \times SU(2)_L \times U(1)_Y
(\times Z_2) \nonumber \\
e. \qquad &&   3_c 1_Q (Z_2) \equiv SU(3)_c \times U(1)_Q (\times Z_2)
\nonumber \\
\end{eqnarray}

\section{Topological defect formation in supersymmetric models}
\label{sec-topo}

In this section, we review the conditions for topological defect
formation during phase transitions in the early universe associated
with the spontaneous symmetry breaking of a group G down to a subgroup
H of G, showing first that the results derived in the non-supersymmetric
case \cite{Kibble} are not affected by the presence of supersymmetry. We
then apply the results to spontaneous symmetry breaking patterns
from supersymmetric SO(10). In a separate section, we study the
formation of hybrid defects, such as monopoles connected by strings
or domain walls bounded by strings, particularly looking at their
cosmological impact \cite{Kib,Vil}.

\subsection{Defects formation in supersymmetric models}

We study here the conditions for defect formation in supersymmetric
models. We show that the conditions for topological defect formation
in non supersymmetric theories \cite{Kibble}, are not affected by the
presence of supersymmetry. We review these conditions with
special application to supersymmetric SO(10).

In non supersymmetric theories, the conditions for topological
defect formation during the spontaneous
symmetry breaking of a non-supersymmetric Lie group G to a
non-supersymmetric Lie group H are well known; they are associated
with the connection of the vacuum manifold ${G \over H}$ \cite{Kibble}.
 Now one may worry
about the non Lie nature of the superalgebra. Fortunately, it has been
shown ~\cite{Srivastava} that the superalgebra is Lie admissible and
that the infinitesimal transformations of the superalgebra can be
exponentiated to obtain a Lie superalgebra. The Lie admissible algebra
is an algebraic covering of the Lie algebra, and it was first
identified by Albert ~\cite{Albert}. It is such a covering that
allows  a Lie admissible infinitesimal behavior while preserving the
global structure of the Lie group. The graded Lie algebra is Lie
admissible and therefore much of the Lie
algebra theory may be extended to it with the appropriate
modification. In particular, a connected (super)Lie group structure
persists ~\cite{Santilli}. Hence, the formation of topological defects in
supersymmetric models
will be the same as in non-supersymmetric ones. Whether or not
supersymmetry is broken at the phase transition will not affect
the conditions under which topological defects form.

The defect formation and stability conditions are therefore as follows
\cite{Kibble}. Consider the spontaneous symmetry breaking of a group G
down to a
subgroup H of G. Topological defects, arising according to the Kibble
mechanism \cite{Kibble} when G breaks down to H, are classified in terms
of the homotopy groups of the vacuum manifold $G\over H$ \cite{Kibble}.
If the fundamental homotopy group $\pi_0 ({G \over H}) \neq I$ is
non trivial, domain walls form when G breaks down to H. If the
first homotopy group $\pi_1({ G \over H}) \neq I$ is non trivial, topological
cosmic strings form. If the second homotopy group $\pi_2({ G \over H})
\neq I$ is non trivial, monopoles form. Note that when
we denote a group G (respectively H), we really mean the
supersymmetric version of this group, and when we write SO(10) we mean
its universal covering group Spin(10) (supersymmetric) which is simply
connected. If the group H breaks later to a subgroup K of H, we have
the following conditions for the stability of the defects formed
when G broke to H. If the fundamental homotopy group
$\pi_0({G\over K})$ is non trivial, the walls are topologically stable,
$\pi_1({G\over K})$ is non trivial, the strings are topologically stable
and if $\pi_2({G\over K})$ is non trivial, the monopoles are
topologically stable down to K. Domain walls, because they are
too heavy, and monopoles, because they are too abundant according
to the Kibble mechanism if present today, would dominate the energy
density of the universe. Hence these defects are in conflict with
the standard cosmology.
 On the other hand, cosmic strings can explain large scale structure,
anisotropies in the Cosmic Background Radiation and part of the baryon
asymmetry of the universe.

Now consider the phase transition associated with the
breaking of SO(10) down to a subgroup G of SO(10), and apply
the above results to this particular case. Since Spin(10) is
connected we have $\pi_2({SO(10) \over G}) = \pi_1(G)$ and $\pi_1({SO(10)
\over G}) = \pi_0(G)$ and therefore the formation of monopoles and strings
during the Grand Unified phase transition is governed by the non
triviality of $\pi_1(G)$ and $\pi_0(G)$ respectively. If G breaks down
later to a subgroup K of G, monopoles formed during the first phase transition
will remain topologically stable after the second phase transition if
$\pi_2({SO(10) \over K}) \neq I$. Strings formed during the first phase
transition will be topologically stable after the next phase transition if
$\pi_1({SO(10) \over K}) \neq I$.

\subsection{Hybrid defects}
\label{sec-topodw}

When we have an intermediate breaking scale, we
can also get mixed defects. There are two kinds of mixed defects that
we can get in supersymmetric SO(10) models; they are monopoles connected
 by strings and domain walls bounded by strings. Their cosmological
evolutions have been studied in a non supersymmetric general case
\cite{Kib,Vil}.

\subsubsection{Monopoles connected by strings}
\label{sec-mncs}

In supersymmetric SO(10) models, we can have monopoles
connected by strings \cite{Vil}. If the first phase
transition leaves an unbroken U(1) symmetry which
later breaks to unity, that is if the breaking pattern proceeds as
\begin{equation}
G \rightarrow H \times U(1)_x \rightarrow H \label{eq:monstr}
\end{equation}
where G and H are both simply connected, then monopoles form
at the first phase transition, and then get connected by strings at the
following one. Indeed, the second homotopy group
$\pi_2({G \over { H\times U(1)}}) = \pi_1(H \times U(1)) = Z$
indicates the formation of monopoles during the first phase transition
in (\ref{eq:monstr}). These monopoles carry a $U(1)_x$ magnetic charge, and are
topologically unstable. Now the first homotopy group
$\pi_1({{H \times U(1)} \over H})$ is also non trivial,
hence cosmic strings form at
the second stage of symmetry breaking in (\ref{eq:monstr}). The strings
connect monopole/antimonopole pairs of the first phase
transition \cite{Vil}. Because the whole system of strings rapidly
decays \cite{Vil},
monopoles connected by strings do not seem to affect the standard
cosmology in any essential way. On the other hand, if the universe
undergoes a period of inflation between the two phase transitions, or
if the phase transition leading to the formation of monopoles is
itself inflationary, then the picture is very different. The decay of the
system of strings  is negligible. If the monopoles are inflated
beyond the horizon,
the strings form according to the Kibble mechanism and their evolution
is that of topologically stable cosmic strings \cite{Vil}. In this class
of scenarios, with inflation and cosmic strings, temperature fluctuations
in the CBR measured by COBE give constraints on the scale of the
phase transition leading to the string formation and on the scalar
coupling constant \cite{next}.

\subsubsection{Walls bounded by strings}

The other kind of topological mixed defect that we can get in SO(10) models
are domain walls connected by strings. A first phase transition leaves
an unbroken discrete symmetry, and cosmic strings form. At a subsequent
phase transition, this discrete symmetry breaks leading to the formations
of domain walls. They are bounded by the strings previously formed.
Specifically, consider a symmetry
breaking pattern of the form
\begin{equation}
G \rightarrow H \times Z_2 \rightarrow H \label{eq:csdw}
\end{equation}
where G and H are both simply connected. The first homotopy group
$\pi_1({G \over { H\times Z_2}}) = \pi_0(H \times Z_2) = Z_2$ thus
$Z_2$-strings form during the first phase transition in (\ref{eq:csdw});
they are topologically unstable. The discrete
$Z_2$ symmetry breaking leads to the formation of domain walls at the
second stage of symmetry breaking bounded by strings
of the first phase transition. Such extended objects have been first
studied by Kibble et al. ~\cite{Kib}.
They have shown that, in the non supersymmetric case, the cosmological
relevance of these mixed objects depends on whether inflation
occurs between the time when strings form and the time when the
symmetry breaking leading to the formation of these walls occurs. The
presence of supersymmetry does not affect the above conclusions.
Following ref. \cite{Kib}, we get the following results. If the
transition leading to the formation of the walls takes place without
supercooling, the walls lose their energy by friction and disappear
in a time $t_d \sim (t_W t_*)^{1 \over 2}$ where $t_W$ is the cosmic
time corresponding to the the scale $T_W$ at which the walls
form and $t_* = {3 \alpha_G \eta_0 \over 32 \pi \eta_3} {M_p^2 \over
 M_G^3}$, where $\eta_3$ is the effective massless degrees of freedom
reflected by the walls and $\eta_0$ is the effective number of degrees
of freedom in the supersymmetric $3_c 2_L 1_Y (Z_2 )$ phase. With $\eta_3 =
33.75$ and $\eta_0 = 228.75$  we find $t_d \sim 10^{-33} - 10^{-36}$ sec
for $T_W \sim 10^{15} - 10^{16}$ GeV and the  corresponding
scale $T_* \sim 10^9 - 10^{12}$ GeV. Therefore these extended objects
do not seem to affect the standard cosmology in any essential way. But
if there is a period of inflation between the two phase transitions,
the strings can be pushed to arbitrarily large scales; the walls form according
to the Kibble mechanism and their evolution is that of topologically
stable walls. The only difference from topologically stable $Z_2$-walls
 is that the walls can now decay by the quantum nucleation of holes
bounded by strings. Hole nucleation however is a tunneling process and
is typically suppressed by a large exponential factor. The corresponding
decay time is much larger than the time at which the walls come to
dominate the universe, thereby upsetting standard cosmology.

\section{Inflation in supersymmetric SO(10) models}
\label{sec-inflation}

Since SO(10) is simply connected and the standard model gauge group
involves an unbroken U(1) symmetry which remains unbroken down to low
energy, all symmetry breaking patterns from supersymmetric SO(10)
down to the standard model automatically involve the formation of
topologically stable monopoles. Even if some monopoles are connected
by strings, a large fraction of them will remain stable down to low
 energy. Hence some mechanism has to be invoked in order to obtain
consistency with the standard cosmology, such as an inflationary
scenario. In this section, we discuss a false vacuum
 hybrid inflationary scenario which is the most natural mechanism
for inflation in global supersymmetric SO(10) models \cite{next}. The
superpotential in the inflaton sector is similar to that studied
in \cite{Shafipot}. We can note first that SO(10) is rank 5, whereas
the standard model gauge group $3_c 2_L 1_Y$ is 4. Hence the rank
of the group has to be lowered from one unit at some stage of the symmetry
breaking. This can be done using a pair of $16 + \overline{16}$ dimensional
Higgs representation, or a pair of $126 + \overline{126}$ dimensional ones if
the $Z_2$ parity is to be kept unbroken, as in models ${\bf 8.}$
to ${\bf 14. }$. We can use a scalar field singlet under SO(10) in order
to force this pair of Higgs to get their VEV on the order of the GUT scale.
The superpotential in this sector will be of the form
\begin{equation}
\alpha {\cal S} \overline{\Phi} \Phi - \mu^2 {\cal S} \label{eq:inflation}
\end{equation}
where $\Phi + \overline{\Phi}$ stand for a pair of $16 + \overline{16}$
dimensional
Higgs representations or a pair of $126 + \overline{126}$ dimensional
Higgs representations, and $\mu \over {\sqrt {\alpha}}$ is assumed to be
the Grand Unified breaking scale. We then identify the scalar field
${\cal S}$ with the inflaton field.

The evolution of the fields is as follows  (a complete discussion
of the potential in a general supersymmetric case is studied in
ref. \cite{Shafipot} and in a specific supersymmetric SO(10) model
is studied in reference \cite{next}). The fields take random
initial values, just
subject to the constraint that the energy density is at the Planck
scale. The inflaton field is distinguished from the other fields
from the fact that the gradient of the GUT potential with respect to the
inflaton field is very small. Therefore the non inflaton fields, except
the $\Phi$ and $\overline{\Phi}$ fields, will roll very
quickly down to their minimum at an approximately fixed value for the
inflaton. Inflation occurs as the inflaton rolls slowly down
the potential. The symmetry breaking implemented with the $\Phi +
\overline{\Phi}$
fields occurs at the end of inflation and associated topological defects
are not inflated away \cite{Shafipot,next}.

\section{SU(5) as intermediate scale}
\label{sec-su5}

We shall describe in this section the symmetry breaking patterns
from supersymmetric SO(10) involving an SU(5) intermediate symmetry.
When the intermediate scale involves SU(5) as a subgroup, say cases
 ${\bf 1}$, ${\bf 2}$ and ${\bf 7}$, the scale $M_G$ has to be $\sim
10^{16}$ GeV, and consequently the scale $M_{GUT}$ is pushed close to
the string compactification scale. $SO(10)$ can break via $SU(5)$ in four
different ways. It can break via $SU(5) \times U(1)_X$, $SU(5)$ , via $SU(5)
\times \widetilde{U(1)}$ and via $SU(5) \times Z_2$, which correspond to models
${\bf 1}$ and ${\bf 8}$, ${\bf 2}$, ${\bf 3}$ and ${\bf 10}$ and ${\bf
9}$respectively.

\subsection{Breaking via $SU(5) \times U(1)_X$}

We consider here two symmetry breaking patterns,
\begin{eqnarray}
&SO(10)& \stackrel{M_{GUT}}{\rightarrow} SU(5) \times U(1)_X\\
&&\stackrel{M_G}{\rightarrow} SU(3)_c \times SU(2)_L \times U(1)_Y (\times Z_2)
\\
&& \stackrel{M_Z}{\rightarrow} SU(3)_c \times U(1)_Q (\times Z_2)
\end{eqnarray}
with and without the $Z_2$ symmetry unbroken down to low energy. The latter
is necessary to preserve large values for the proton lifetime and to
stabilize the LSP. It can arise only if a pair of $126 + \overline{126}$
dimensional Higgs representations are used to lower the rank of the group,
and hence must be part of the standard model gauge group in order to give
large Majorana mass to the right-handed neutrino.

The $U(1)_X$ commutes with SU(5). The X and Y directions are orthogonal
to each other, and thus the $U(1)_X$ symmetry breaks down to unity at
$M_G$ (or to $Z_2$ if a pair of $126 + \overline{126}$ Higgs fields are used to
break $SU(5) \times U(1)_X$). This feature is going to affect the
formation of topological defects.

The first homotopy group $\pi_1 (SU(5) \times U(1)_X) = Z$ is non trivial
and thus topological monopoles form when SO(10) breaks. They have a mass
$M_m
\geq 5 \times 10^{17}$ GeV. At the following phase transition the $U(1)_X$
symmetry breaks to unity (to $Z_2$) and hence cosmic strings
($Z_2$-strings) form. They  connect monopole-antimonopole pairs
previously formed (see section \ref{sec-mncs}). They have a mass
per unit length $\sim 10^{32} \: GeV^2$.

When $SU(5) \times U(1)_X$ breaks down to $3_c 2_L 1_Y (Z_2)$ new lighter
monopoles form. Indeed, since $U(1)_X$ breaks down to unity (to $Z_2$) we
consider the second homotopy group $\pi_2 ({ SU(5) \over 3_c 2_L 1_Y})$
to look for monopoles formations at $M_G$. Hence topologically stable
monopoles form. They have a mass $M_m \sim 10^{17}$ GeV. They are topologically
stable. Their topological charge may change from Y to Q.

Since monopoles form at both phase transitions and since the lighter ones
are topologically stable, the inflationary scenario, as in section
\ref{sec-inflation}, is unable to solve the monopole problem. Hence these two
models are inconsistent with observations.

\subsection{Breaking via SU(5)}
\label{sec-alone}

Here, SO(10) breaks down to the standard model with intermediate SU(5) symmetry
alone. In this case, there is no interest in going to a larger Grand Unified
group. The breaking scheme is
\begin{eqnarray}
&SO(10)& \stackrel{M_{GUT}}{\rightarrow} SU(5) \\
&&\stackrel{M_G}{\rightarrow} SU(3)_c \times SU(2)_L \times U(1)_Y \\
&& \stackrel{M_Z}{\rightarrow} SU(3)_c \times U(1)_Q
\end{eqnarray}
which is that of model ${\bf 1}$. Since $SO(10)$ and SU(5) are both simply
connected, no topological defects form during the first stage of symmetry
breaking.

The second homotopy group $\pi_2({SU(5) \over 3_c 2_L 1_Y}) = Z$
hence topological monopoles form when $SU(5)$ breaks down to the
standard model.
The monopoles carry Y topological charge. The second homotopy
group $\pi_2({SU(5) \over 3_c 1_Q}) = Z$ which shows that the monopoles
are topologically stable. They have a mass $M_m \sim \: 10^{17}$ GeV.
Their topological charge may change from Y to Q.

Since the rank of SO(10) is 5 and the rank of $SU(5)$ is 4, if we use an
inflationary scenario as described in sec. \ref{sec-inflation} to solve
the monopole problem, the inflaton field will
couple to a pair of $16 + \overline{16}$ Higgs fields representations which
will be used used to break SO(10).
The monopoles described above will form at the end of inflation,
and their density will be high enough to dominate the universe. Hence
this model is in conflict with the standard cosmology. It is also inconsistent
with the actual data on the proton lifetime.

\subsection{Breaking via $SU(5) \times \widetilde{U(1)}$}

More interesting is the breaking via flipped $SU(5)$
\begin{eqnarray}
&SO(10) &\stackrel{M_{GUT}}{\rightarrow} SU(5) \times \widetilde{U(1)} \\
&&\stackrel{M_G}{\rightarrow} SU(3)_c \times SU(2)_L \times U(1)_Y  \\
&& \stackrel{M_Z}{\rightarrow} SU(3)_c \times U(1)_Q
\end{eqnarray}
Note that with flipped SU(5), rather than using SO(10) for the Grand
Unified gauge group, the monopole problem is avoided \cite{nano}.
The $\widetilde{U(1)}$ contains part of the electromagnetic gauge
group $U(1)_Q$. The above symmetry breaking can only be implemented
in supergravity SO(10) models \cite{nano}.

The first homotopy group $\pi_1(SU(5) \times \widetilde{U(1))} = Z$ and
therefore the
first phase transition leads to the formation of topological monopoles
when SO(10) breaks. Furthermore, since $\pi_1(3_c 2_L 1_Y) =
\pi_1(3_c 1_Q) = Z$ and $\widetilde{U(1)}$ contains part of the $U(1)_Y$ and
$U(1)_Q$ symmetries, these monopoles are topologically stable. They have a
mass $M_m \geq 5 \times 10^{17}$ GeV. They carry $B\! -\! L$, and
their topological charge may change to Y and then to Q. Embedded cosmic
strings form after the second stage of symmetry breaking \cite{Nathan}.

We should be able to cure the monopole problem with an hybrid inflationary
scenario for supergravity models. Indeed, since
the rank of $SU(5) \times \widetilde{U(1)}$ is 5, the inflaton field can couple
to the Higgs needed to break $SU(5) \times \widetilde{U(1)}$, and embedded
strings
will form at the end of inflation.
Hence from a defects point of view the model is interesting, but appears
to be inconsistent  with the actual data for proton lifetime
\cite{Martin} and does not provide
any Majorana mass for the right-handed neutrino. The latter problems are
solved if we break $SU(5) \times
\widetilde{U(1)}$ down to $3_c 2_L 1_Y Z_2$. In that case, a $126 +
\overline{126}$ dimensional Higgs representation is used to break
$SU(5) \times \widetilde{U(1)}$. Since the
first homotopy groups $\pi_1( {SU(5) \times \widetilde{U(1)}
\over 3_c 2_L 1_Y Z_2})
= Z_2$ and $\pi_1( {SU(5) \times U(1) \over 3_c 1_Q Z_2})
= Z_2$, topologically stable $Z_2$-strings also form. They have a mass
per unit length $\sim 10^{32} \: GeV^2$.

\subsection{Breaking via $SU(5) \times Z_2$}

We consider here the breaking of SO(10) via SU(5) with added parity. The
symmetry breaking is
\begin{eqnarray}
&SO(10) &\stackrel{M_{GUT}}{\rightarrow} SU(5) \times Z_2 \\
&&\stackrel{M_G}{\rightarrow} SU(3)_c \times SU(2)_L \times U(1)_Y \times Z_2
\\
&& \stackrel{M_Z}{\rightarrow} SU(3)_c \times U(1)_Q \times Z_2
\end{eqnarray}
 where the unbroken $Z_2$ symmetry is a subgroup of the $Z_4$ centre of
SO(10). It plays the role of matter parity. It preserves large values for
the proton lifetime and stabilizes the LSP, thus the model is consistent
with the actual data on proton decay and provide a good hot dark
matter candidate.

Now the fundamental homotopy group $\pi_0(SU(5) \times Z_2) = Z_2$ and
therefore $Z_2$ cosmic strings form
during the first phase transition. They have a mass per
unit length $10^{38} GeV^2 \geq \mu \geq 10^{32} \: GeV^2$. Since the $Z_2$
symmetry is kept unbroken down to low energy, these strings
remain topologically stable. They have been widely
studied in the non supersymmetric case \cite{so10cs}.

As in section \ref{sec-alone}, it is clear that topologically stable
monopoles form during the second phase transition with mass $M_m
\sim 10^{17}$ GeV. Hence as in section \ref{sec-alone}, the model
is in contradiction with observations.

\vspace{1cm}

We conclude that the only symmetry breaking pattern from SO(10) down to the
standard model with intermediate SU(5) symmetry consistent with observations,
is
\begin{equation}
 SO(10) \rightarrow SU(5) \times \widetilde{U(1)} \rightarrow 3_c 2_L 1_Y Z_2
\rightarrow 3_c 1_Q Z_2
\end{equation}
where the $Z_2$ symmetry must be kept unbroken in order to
preserve large values for the proton lifetime. The above symmetry
breaking can only be implemented in supergravity models.

\section{Patterns with a Left-Right intermediate scale}
%\label{sec-leftr}

In this section we study the symmetry breaking patterns from
supersymmetric SO(10) down to the standard model involving an
$SU(2)_L \times SU(2)_R$ intermediate symmetry. These are the
symmetry breaking patterns with intermediate $4_c 2_L 2_R (Z_2)$
or $3_c 2_L 2_R 1_{B-L}$ symmetry groups. We show that these models, due
the unbroken $SU(2)_L \times SU(2)_R$ symmetry share a property,
which can make them cosmologically irrelevant, depending on the Higgs field
chosen to implement the symmetry breaking.  We then give a full study of the
formation of the topological defects in each model.

\subsection{Domain walls in left-right models}
\label{sec-DW}

We study here a property shared by the symmetry breaking schemes from
SO(10) down to the standard model, with or without unbroken parity $Z_2$,
\begin{equation}
SO(10)  \stackrel{M_{GUT}}{\rightarrow} G  \stackrel{M_{G}}{\rightarrow}
3_c 2_L 1_Y (Z_2) \label{eq:LR1}
\end{equation}
where G is either $4_c 2_L 2_R$ or $3_c 2_L 2_R 1_{B-L}$. In these models,
the intermediate scale involves an unbroken $SU(2)_L \times SU(2)_R$
symmetry, and consequently the intermediate symmetry group can be
invariant under the charge conjugation operator, depending on the
Higgs multiplet chosen to break SO(10). The latter leaves an unbroken
discrete $Z_2^C$ symmetry which breaks at the following phase transition.
In this case, the general symmetry breaking scheme given in equation
(\ref{eq:LR1}) should really be written as
\begin{equation}
SO(10) \stackrel{M_{GUT}}{\rightarrow} G
\times Z_2^C \stackrel{M_G}{\rightarrow} SM (\times Z_2)  \label{eq:LR3} \: .
\end{equation}
If $G = 4_c 2_L 2_R$, the discrete $Z_2^c$ symmetry appears if the Higgs
used to break SO(10) is a single 54 dimensional representation \cite{54}.
If $G= 3_c 2_L 2_R 1_{B-L}$ the $Z_2^c$ symmetry appears if a single 210
dimensional Higgs representation is used, with appropriate parameter
range in the Higgs potential \cite{210}. The appearance of the discrete
$Z_2^c$ symmetry leads to a cosmological problem \cite{Kib}. Indeed,
since Spin(10) is simply connected, $\pi_1({{SO(10)} \over {G \times
Z_2^c}}) = \pi_0(G \times Z_2^c ) = Z_2$ and therefore $Z_2$ strings
form during the first phase transition associated with the breaking of
SO(10). They have a mass per unit length $\sim 10^{32} - 10^{34} \: GeV^2$.
When
the discrete $Z_2^C$ symmetry breaks, domain walls form bounded by the
strings of the previous phase transition. Some closed walls can also
form. As shown in section \ref{sec-topodw}, these domain walls do not
affect the standard cosmology in any essential way. On the other
hand, if a period of inflation occurs between the two phase transition, or
if the phase transition leading to the walls formation is itself
inflationary, then the evolution of the walls is that of topologically
stable $Z_2$ walls. They dominate the universe, destroying the
standard cosmology.

\subsection{Breaking via $4_c 2_L 2_R$}

We now consider the symmetry breaking of SO(10) via the Pati-Salam gauge group
$4_c 2_L 2_R$ subgroup of SO(10) which later breaks down to the
standard model gauge group with or without matter parity
\begin{eqnarray}
&SO(10)& \stackrel{M_{GUT}}{\rightarrow} SU(4)_c \times SU(2)_L \times
 SU(2)_R \\
&&\stackrel{M_G}{\rightarrow} SU(3)_c \times SU(2)_L \times U(1)_Y
(\times Z_2) \\
&&\stackrel{M_Z}{\rightarrow} SU(3)_c \times U(1)_Q  (\times Z_2) \label{eq:PS}
\end{eqnarray}
with supersymmetry broken at $\simeq 10^3$ GeV and the scales $M_{GUT}$
and $M_G$ respectively satisfy $M_{pl} \geq M_{GUT} \geq 10^{16}$ GeV
and $M_G \sim 10^{15} - 10^{16}$ GeV.
The discrete $Z_2$ symmetry is kept unbroken if we use a pair of
$126 + \overline{126}$-Higgs dimensional representation to break $4_c 2_l 2_R$,
and is broken if we use a pair of $16 + \overline{16}$ dimensional
Higgs. The unbroken $Z_2$ symmetry plays the role of matter parity,
preserving large values for the proton lifetime and stabilizing the LSP.
Hence only the model with unbroken $Z_2$ at low energy is consistent
with the actual value for proton lifetime.

If a single 54 dimensional Higgs representation is used to break SO(10),
equation (\ref{eq:PS}) should really be written as \cite{Kib}
\begin{eqnarray}
&Spin(10) &\stackrel{M_{GUT}}{\rightarrow} ({(Spin(6) \times Spin(4))
\over Z_2 }) \times Z_2^C \label{eq:PS2}\\
&&\stackrel{M_G}{\rightarrow} SU(3)_c \times SU(2)_L \times U(1)_Y
(\times Z_2) \\
&&\stackrel{M_Z}{\rightarrow} SU(3)_c \times U(1)_Q (\times Z_2)
\end{eqnarray}
where we have explicitly shown the hidden symmetry. A $Z_2$ symmetry has
to be factored out in equation (\ref{eq:PS2}) since $Spin(6)$ and $Spin(4)$
have a non trivial intersection. The overall $Z_2^c$  is generated by the
charge conjugation operator; it is unrelated to the previous $Z_2$ one.
Subsequently, the $Z_2^c$ discrete symmetry is broken. If a pair of Higgs
in the ${126 + \overline{126}}$ representation are used to break $4_c 2_l 2_R$,
then a new $Z_2$ symmetry emerges, as described above; it is unrelated
to the previous ones. The standard model gauge group is broken with a
Higgs in the 10 dimensional representation of SO(10).

If a single {210}-Higgs multiplet is used to break $4_c 2_l 2_R$, with
appropriate range in the parameters of the Higgs potential, the $Z_2^C$
does not appear \cite{210}.

\subsubsection{Monopoles}
\label{sec-1}

The non trivial intersection of $Spin(6)$ and $Spin(4)$ leads to the
production of superheavy monopoles \cite{Vil} when SO(10) breaks to
$4_c 2_L 2_R$. These monopoles are superheavy with a mass $M_m \geq
10^{17}$ GeV. They are topologically unstable.

Since the second homotopy group $\pi_2({ 4_c 2_L 2_R \over  3_c 2_L 1_Y
(Z_2)}) = Z$ is non trivial, new monopoles form when $4_c 2_L 2_R $ breaks
down to the standard model gauge group. They are unrelated to the previous
monopoles. Furthermore, since the second homotopy group $\pi_2({ 4_c 2_L 2_R
\over  3_c 1_Q (Z_2)}) = Z$ is also non-trivial, these lighter monopoles
 are topologically stable. They have a mass $M_m \sim 10^{16} - 10^{17}$ GeV.
These monopoles form according to the Kibble mechanism, and
their density is such that, if present today, they would dominate the
energy density of the universe.

\subsubsection{Domain walls}
\label{sec-ps}

If a 54 dimensional Higgs representation is used to break SO(10)
down to $4_C 2_L 2_R$, the symmetry breaking is given by
equation (\ref{eq:PS2}) which is of the form of equation (\ref{eq:LR3})
with $G= 4_c 2_L 2_R$, so that a discrete $Z_2^c$ symmetry emerges at the
intermediate scale. Thus, as shown in section \ref{sec-DW}, $Z_2$-strings
form during the first phase transition. (They are unrelated
to any of the monopole just discussed above.) During the second stage of
symmetry breaking, this $Z_2^C$ breaks leading to the formation
of domain walls which connect the strings previously formed.
These walls bounded by strings do not affect the standard cosmology
in any essential way. But if there is a period of inflation before
the phase transition leading to the walls formation takes place (see
section \ref{sec-topodw}), the walls would dominate the energy density
of the universe, leading to a cosmological catastrophe.

\subsubsection{Cosmic strings}

Now we consider the models where $4_c 2_L 2_R$ breaks down to the standard
model gauge group with added $Z_2$ parity, as in model ${\bf 8}$. Then a
new $Z_2$ symmetry emerges at $M_G$, which is unrelated to the previous ones.
Since $\pi_1({4_c 2_L 2_R  \over 3_c 2_L 1_Y Z_2}) = Z_2$  $Z_2$-strings form
when $4_c 2_L 2_R$ breaks. They have a mass per unit length $\mu \sim
10^{30} - 10^{32} \: GeV^2$. Since the $Z_2$ symmetry is then kept unbroken
down to low energy, we break the standard model gauge group with a Higgs
10-plets. The strings are topologically stable down to low energy.

Density perturbations in the early universe and temperature fluctuations
in the CBR generated by these strings could be computed.

\subsubsection{Solving the monopole problem}

In order to solve the monopole problem, we use
an hybrid inflationary scenario, as discussed in section
\ref{sec-inflation}. The rank of both $4_c 2_L 2_R$ and $4_c 2_L 2_R Z_2$
is four. Therefore the inflaton field will couple to a pair of
Higgs field which will break $4_c 2_L 2_R$. The primordial monopoles formed
when SO(10) breaks are diluted by the inflation. But then lighter monopoles
form at the end of inflation when $4_c 2_L 2_R$ breaks, which are
topologically stable. In the case
of unbroken $Z_2$ parity cosmic strings also form. Monopole creation at this
later stage make the model inconsistent with observations.

If SO(10) is broken with a 54 dimensional Higgs representation, domain
walls will form through the Kibble mechanism at the end of inflation,
which will dominate the universe, as shown in section \ref{sec-DW}, hence
leading to a cosmological catastrophe.

We conclude that the model is cosmologically inconsistent with observations.
It is inconsistent whether or not the discrete
$Z_2^C$ symmetry is unbroken at the intermediate scale.

\subsection{Breaking via $3_c 2_L 2_R 1_{B-L}$}

We can break via $3_c 2_L 2_R 1_{B-L}$ and then down to the standard model
with or without the discrete $Z_2$ symmetry preserved at low energy
\begin{eqnarray}
& SO(10) & \stackrel{M_{GUT}}{\rightarrow}SU(3)_c \times SU(2)_L
\times  SU(2)_R \times U(1)_{B\! -\! L} \label{eq:above}
\\
&&\stackrel{M_G}{\rightarrow} SU(3)_c \times SU(2)_L \times U(1)_Y
(\times Z_2) \\
&& \stackrel{M_Z}{\rightarrow} SU(3)_c \times U(1)_Q (\times Z_2)
\end{eqnarray}
The $Z_2$ symmetry, which can be kept unbroken down to low energy if only
safe representations are used to implement the symmetry breaking, plays
the role of matter parity. It preserves large values for the proton lifetime.
Hence only models with unbroken $Z_2$ parity at low energy are consistent
with the actual values of proton decay. If SO(10) is broken with a single
210-Higgs multiplet, with the appropriate range of the parameters in the
Higgs potential \cite{210}, then there appears a discrete $Z_2^c$ symmetry at
the intermediate scale which is generated by the charge conjugation operator,
and the symmetry breaking really is
\begin{eqnarray}
& SO(10) & \stackrel{M_{GUT}}{\rightarrow}SU(3)_c \times SU(2)_L \times
 SU(2)_R \times U(1)_{B\! -\! L} \times Z_2^c \label{eq:LR2}\\
&&\stackrel{M_G}{\rightarrow} SU(3)_c \times SU(2)_L \times U(1)_Y
(\times Z_2)\label{eq:21}\\
&& \stackrel{M_Z}{\rightarrow} SU(3)_c \times U(1)_Q (\times Z_2)\label{eq:22}
\: .
\end{eqnarray}
The $Z_2^c$ is unrelated to the $Z_2$ symmetry which can be added to the
standard model gauge group in equations (\ref{eq:21}) and (\ref{eq:22}).
If one uses a combination of a 45 dimensional Higgs representation with a 54
dimensional one to break SO(10),
then the symmetry breaking is that of equation (\ref{eq:above}), and no
discrete symmetry appears as in (\ref{eq:LR2}) \cite{45}. The rest of
the symmetry
breaking is implemented with a pair of $16 + \overline{16}$-Higgs multiplets
or with a pair of $126 + \overline{126}$-Higgs multiplets if matter parity
is preserved at low energy. $3_c 2_L 1_Y$ is broken with a 10-Higgs
multiplet.

\subsubsection{Monopoles}

The first homotopy groups $\pi_1(3_c 2_L 2_R 1_{B-L}) = Z$, $\pi_1(3_c 2_L
1_{Y}) = Z$ and $\pi_1(3_c 1_{Q}) = Z$, showing that topologically stable
monopoles are produced during
the first phase transition from SO(10) down to $3_c 2_L
2_R 1_{B-L}$. They have a mass $M_m \geq 10^{17}$
GeV. These monopoles are in conflict with cosmological observations.

\subsubsection{Domain walls}

If SO(10) is broken with a single 210 dimensional Higgs representation
then the symmetry breaking is that of equation (\ref{eq:LR}). Hence, as
in the breaking pattern (\ref{eq:PS2}), the appearance of the
discrete $Z_2^c$ symmetry leads to the
formation of non-stable cosmic strings during the first symmetry breaking
 and to the formation of domain walls in the breaking of  $3_c 2_L
2_R 1_{B-L}$ down to the standard model gauge group. The cosmological
relevance of these walls bounded by strings depends upon the presence of an
inflationary epoch before the phase transition leading to the walls formation
has taken place, see sec. \ref{sec-DW}.

\subsubsection{Embedded Defects}

In these models with intermediate $3_c 2_L 2_R 1_{B-L}$ symmetry, the
breaking schemes are equivalent to
\begin{equation}
SO(10) \stackrel{M_{GUT}}{\rightarrow} G \times  SU(2)_R \times
U(1)_{B\! -\! L} \stackrel{M_G}{\rightarrow} G \times U(1)_Y (\times Z_2)
\stackrel{M_Z}{\rightarrow} 3_c 1_Q (Z_2)
\end{equation}
where $G = SU(3)_c \times SU(2)_L$. In direct analogy with electroweak
strings \cite{Tanmay}, it is easy to see that embedded defects form
during the second stage
of symmetry breaking. They have a mass per unit length $\mu \sim
10^{30} - 10^{32} \: GeV^2$. The stability conditions for these
strings can be computed. If these strings are dynamically stable,
they may generate density perturbations
in the early universe and temperature anisotropy in the microwave
background.

\subsubsection{Cosmic Strings}

Consider the model where $3_c 2_L 2_R 1_{B-L}$ breaks down to
$3_c 2_L 1_Y Z_2$. The first homotopy group $\pi_1({3_c
2_L 2_R 1_{B-L} \over 3_c
2_L 1_Y Z_2}) = Z_2$ is non trivial which shows the
formation of topological $Z_2$ strings. Since the $Z_2$ parity symmetry is
kept unbroken down to low energy, the strings are
topologically stable. They have a mass per unit length $\mu
\sim 10^{30}-10^{32} \: GeV^2$. These strings will generate density
perturbations in the early universe and temperature anisotropy in the
microwave background.

\subsubsection{Solving the monopole problem}

One can use an inflationary scenario as described in section
(\ref{sec-inflation})
to dilute the monopoles formed at $M_{GUT}$. Since the rank of $3_c
2_L 2_R 1_{B-L} (Z_2^c)$ is four, the inflaton field will couple to a
pair of $16 + \overline{16}$ or
$126 + \overline{126}$ which will break $3_c 2_L 2_R 1_{B-L}$, (see section
\ref{sec-inflation}). Cosmic strings (if unbroken $Z_2$ symmetry at low energy)
and/or domain walls (if unbroken $Z_2^c$ symmetry at the intermediate
scale) will form at the end of inflation. As shown in section \ref{sec-DW}
the presence of this inflationary epoch between the two phase transitions at
$M_{GUT}$ and $M_G$ respectively would make the walls dominate the energy
density of the universe, (see sec. \ref{sec-DW}). Now the unbroken $Z_2$
symmetry is necessary to preserve large values for the proton life time, hence
the only symmetry breaking pattern consistent with cosmology with intermediate
$3_c 2_L 2_R 1_{B-L}$ symmetry is
\begin{eqnarray}
& SO(10) & \stackrel{<45> + <54>}{\rightarrow}SU(3)_c \times SU(2)_L \times
 SU(2)_R \times U(1)_{B\! -\! L} \label{eq:LR}\\
&&\stackrel{<126> + <\overline{126}>}{\rightarrow} SU(3)_c \times SU(2)_L
\times U(1)_Y
\times Z_2\\
&& \stackrel{<10>}{\rightarrow} SU(3)_c \times U(1)_Q \times Z_2
\end{eqnarray}
where SO(10) is broken with a combination of a 45 dimensional Higgs
representation and 54 dimensional one,  $3_c
2_L 2_R 1_{B-L}$ is broken with pair of $126 + \overline{126}$ dimensional
Higgs representation and  $3_c 2_L 2_Y Z_2$ is broken with a 10 Higgs
multiplet.

\section{Breaking via $3_c 2_L 1_R 1_{B-L}$}
\label{sec-three}

We shall consider first the symmetry breaking with intermediate  $3_c 2_L 1_R
1_{B-L}$ without conserved matter parity at low energy
\begin{eqnarray}
& SO(10)& \stackrel{M_{GUT}}{\rightarrow} SU(3)_c \times SU(2)_L \times U(1)_R
\times U(1)_{B\! -\! L} \\
&&\stackrel{M_G}{\rightarrow} SU(3)_c \times SU(2)_L \times U(1)_Y\\
&& \stackrel{M_Z}{\rightarrow} SU(3)_c \times U(1)_Q
\end{eqnarray}
The first homotopy group $\pi_1(3_c 2_L 1_R 1_{B\! -\! L}) = Z + Z$ and
therefore
topological monopoles form during the first phase transition
from supersymmetric SO(10) down to $3_c 2_L 1_R 1_{B-L}$. These monopoles
carry $R$ and $B\! -\! L$, and have a mass $M_m \geq 10^{16} -
10^{17}$ GeV. Now $\pi_1(3_c 2_L 1_Y)$
and  $\pi_1(3_c \times 1_Q)$ are both non trivial
and hence, from an homotopy point of view, the monopoles are
topologically stable. But as we are going to show below, some of these
monopoles are indeed topologically stable, but some others will decay.
During the second phase transition, the formation of strings is
governed by the first homotopy group $\pi_1({3_c 2_L
1_R 1_{B\! -\! L} \over 3_c 2_L 1_Y}) = Z $ showing the formation of cosmic
strings during the
second phase transition. These are associated with the breaking of
$U(1)_R \times U(1)_{B\! -\! L}$ down to $U(1)_Y$ where the unbroken
$U(1)_R \times U(1)_{B\! -\! L}$ symmetry in the first stage of
symmetry breaking is responsible for the formation of monopoles. Now
the weak hypercharge ${Y \over 2}$ is a linear combination of $B\! -\! L$
and
$R$, ${Y \over 2} = ({{B\! -\! L} \over 2}  + R)$. Therefore primordial
monopoles with topological charge ${{B\! -\! L} \over 2} -  R  \neq 0$
get connected by
the strings at the second stage of symmetry breaking. Some infinite
and closed strings can also form. These cosmic strings are
topologically unstable. They can break producing
monopole-antimonopole pairs at the free ends. The
monopole/antimonopole pairs connected by strings annihilate in less
than a Hubble time and could produce the observed baryon asymmetry of
the universe \cite{next}. Other monopoles formed during the first phase
transition do not
get connected by strings and remain stable down to low energy.

The monopole problem can be solved with an inflationary scenario
as described in section \ref{sec-inflation}. Since the rank of
$3_c 2_L 1_R 1_{B-L}$ is 5, the inflaton field will couple to the
Higgs mediating the second phase transition associated with the breaking of
$3_c 2_L 1_R 1_{B-L}$. The monopoles
can be pushed beyond the present horizon, and the monopole problem
solved. Furthermore, since all the monopoles are inflated away, the
string decay probability is negligible and the evolution of strings is
identical to that of topologically stable strings. We therefore have a
very interesting breaking scheme, where monopoles are created during a
first transition, inflated away before cosmic strings which can
explain galaxy formation, form.

This model where $3_c 2_L 1_R 1_{B\! -\! L}$ breaks down to the standard model
without matter parity is in conflict with
the actual data for proton lifetime. The solution to this problem is
therefore that the intermediate subgroup break down to $3_c
2_L 1_Y Z_2$ as in model ${\bf 13}$. In this case, topologically
stable $Z_2$-strings will form during the second phase
transition. They have a mass per unit length $\mu \sim 10^{30} -
10^{32} \: GeV^2$.
This interesting model with inflation and cosmic strings is studied in detail
elsewhere \cite{next}.

\section{Breaking directly to the standard model}
\label{sec-direct}

Supersymmetric SO(10) can break directly down to the standard model as
in model ${\bf 7}$
\begin{equation}
SO(10) \stackrel{M_{GUT}}{\rightarrow}  SU(3)_c \times SU(2)_L \times U(1)_Y
\stackrel{M_Z}{\rightarrow} SU(3)_c \times U(1)_Q \label{eq:d2}
\end{equation}
or as in model ${\bf 14}$
\begin{equation}
SO(10) \stackrel{M_{GUT}}{\rightarrow} SU(3)_c \times SU(2)_L \times U(1)_Y
\times Z_2
 \stackrel{M_Z}{\rightarrow} SU(3)_c \times U(1)_Q \times Z_2 \label{eq:d1}
\end{equation}
with (\ref{eq:d1}) or without (\ref{eq:d2}) the $Z_2$ symmetry, subgroup of
the $Z_4$ centre of SO(10), unbroken down to low energy. The latter plays
the role of matter parity, giving large values for the proton lifetime
and stabilizing the LSP. The symmetry breaking occurs at $M_{GUT} \simeq
2 \times 10^{16} GeV$. The scenario without the unbroken $Z_2$ symmetry
(\ref{eq:d2}) is not, with the present data for proton decay, relevant
phenomenologically. The $Z_2$ symmetry is also necessary for stabilizing
the LSP and to
provide a good cold dark matter candidate.

In model (\ref{eq:d1}), the $Z_2$ symmetry remains unbroken
down to low energy preserving large values for the proton
lifetime. Furthermore, the first homotopy group $\pi_1( {{SO(10)} \over
{3_c 2_L 1_Y Z_2}}) = \pi_0 ( 3_c 2_L 1_Y Z_2) = Z_2$ and therefore
cosmic strings form when SO(10) breaks. They are
associated with the unbroken $Z_2$ symmetry and since the latter
remains unbroken down to low energy, the strings are topologically
stable down to low energy. They have a mass per unit length $\mu \sim
10^{32} \: GeV^2$. The latter could account for the density perturbations
produced in the early universe which lead to galaxy formation and to
temperature fluctuations in the CMBR.

Again, due to the unbroken $U(1)_Y$
symmetry, monopoles form at the Grand Unified phase transition. They
carry $Y$ topological charge and are topologically stable down to low
energy. Their topological charge may change from $Y$ to $Q$.

Since monopoles form in both models, the potential conflict with the
standard big bang cosmology is again not avoided. Nevertheless, in
model (11), if the Higgs field leading to monopole production takes
its VEV before inflation ends and the latter ends
before the Higgs leading to cosmic string formation acquires its VEV
then we are left with a very attractive scenario.

Unfortunately, it does not seem possible to achieve this. If one attempts
to inflate away the monopoles with a superpotential of the form given
in section \ref{sec-inflation}, an intermediate  scale is introduced.
Thus, one is either left with the monopole problem in cosmology or
loses the simplicity of this breaking scheme.

\section{Conclusions}
\label{sec-concl}

The aim of this paper is to constrain supersymmetric SO(10) models
which lead to the formation of topological defects through cosmological
considerations. The main reason for
considering supersymmetric versions of the Grand Unified gauge group
SO(10) rather than non-supersymmetric ones, is to predict
the measured values of $\sin^2 \theta_w$ and the gauge coupling
constants merging in a single point at $\sim 2 \times 10^{16}$ GeV. Spontaneous
Symmetry Breaking patterns from supersymmetric SO(10) down to the
standard model differ from
non-supersymmetric ones firstly in the scale of $B-L$
symmetry breaking and secondly in the ways of breaking from SO(10) down to
the standard model. For non-supersymmetric models the scale of $B-L$
breaking has to be anywhere between $10^{10}$ and $10^{13.5}$ GeV
whereas it is $10^{15}$ to $10^{16}$ GeV in supersymmetric models.
Furthermore, in the supersymmetric case, we can break directly down to
the standard model without any intermediate breaking scale, and not
more than one intermediate scale is expected. We have given a systematic
analysis of topological defects formation
and their cosmological implications in each model. We found that the
rules for topological defect formation are not affected by the
presence of supersymmetry and since SO(10) is simply connected and the
standard model gauge group involves an unbroken U(1) symmetry, all
SSB patterns from supersymmetric SO(10) down to the standard model
involve automatically the formation of topologically stable monopoles.
In tables 1, 2, 3 and 4 we give a summary of all the defects formed in
each model. In the models where $Z_2$-walls arise at the second phase
transition, we have in fact hybrid defects. The walls are bounded by the
$Z_2$-strings previously formed and are unstable. In order to solve
the monopole problem, we propose an hybrid inflationary scenario
\cite{Cop,Shafipot,next} which arise in supersymmetric SO(10) models
without imposing any external symmetry and without imposing any external
field \cite{next}. The inflationary scenario can cure the monopole problem, but
then stabilizes the $Z_2$ walls previously discussed. Hence these
cases lead to another cosmological problem. Imposing also that the models
satisfy the actual data on the proton lifetime,
we found that there are only two spontaneous symmetry breaking
patterns consistent with cosmological considerations. Breaking directly
to the standard model at first sight seems attractive. Unfortunately,
one is unable to inflate away the monopoles without the introduction of
an intermediate scale. The only breaking schemes consistent with
cosmology  correspond to the intermediate symmetry groups
$3_C 2_L 2_R 1_{B-L}$, where SO(10) is broken
with a combination of a 45 dimensional Higgs representation and a
54 dimensional one, and $3_C 2_L 1_R 1_{B-L}$. These intermediate
symmetry groups
must later break down to the standard model with unbroken matter parity;
the symmetry breaking must be implemented with only Higgs fields in 'safe'
representations \cite{Martin}, hence the rank of the group must be lowered
with a pair of Higgs in the
$126 + \overline{126}$ dimensional representation, and the standard model
gauge group broken with a 10 dimensional one. The model with
intermediate $3_C 2_L 1_R 1_{B-L}$, inflation and cosmic strings, is
studied in detail elsewhere \cite{next}. In supergravity SO(10) models,
the breaking of SO(10) via flipped SU(5) is also possible.

\section*{Acknowledgements}

We would like to thank Q. Shafi for useful discussions. We would also
like to thank E. Copeland, E. Kiritsis, A. Liddle and
D. Lyth. Finally, we would like to thank Newnham College and PPARC for
financial support, and the Isaac Newton Institute for Mathematical
Sciences for hospitality while this work was in progress.

\newpage
\appendix
\section{}
\begin{tabular}{|l|l|l|l|}
\hline
G & $SO(10) \rightarrow G$ & $G \rightarrow 3_c 2_L 1_Y$ &  cosmological
problems \\ \hline
$SU(5) \times U(1)_X$ & monopoles-1 &    \begin{tabular} {l}
 monopoles-2  \\
+ strings
\end{tabular}&
\begin{tabular} {l}
 monopoles-2  \\
+ proton lifetime   ($Z_2$ broken)
\end{tabular}\\ \hline
SU(5) & no defects & monopoles & \begin{tabular} {l}
 monopoles \\
+ proton lifetime ($Z_2$ broken)
\end{tabular}\\ \hline
$SU(5) \times \widetilde{U(1)}$ & monopoles & embedded strings &
\begin{tabular} {l}
proton lifetime   ($Z_2$ broken)
\end{tabular} \\ \hline
$4_c 2_L 2_R$ &  monopoles-1 &  monopoles-2 & \begin{tabular} {l}
monopoles-2  \\
+ proton lifetime   ($Z_2$ broken)
\end{tabular}\\ \hline
$4_c 2_L 2_R Z^c_2$ & \begin{tabular} {l}
  monopoles-1 \\
+ $Z_2$-strings
\end{tabular}&  \begin{tabular} {l}
monopoles-2 \\
$Z_2$-walls \end{tabular} & \begin{tabular} {l}
$Z_2$-walls and monopoles-2  \\
+ proton lifetime ($Z_2$ broken)
\end{tabular}\\ \hline
$3_c 2_L 2_R 1_{B-L}$ & monopoles & embedded strings & \begin{tabular} {l}
proton lifetime  ($Z_2$ broken)
\end{tabular}
\\ \hline
$3_c 2_L 2_R 1_{B-L} Z^c_2 $ & \begin{tabular} {l}
monopoles \\
+$Z_2$-strings
\end{tabular}& \begin{tabular} {l}
embedded strings \\
+ $Z_2$-walls
\end{tabular} & \begin{tabular} {l}
 $Z_2$-walls \\
+ proton lifetime  ($Z_2$ broken)
\end{tabular} \\ \hline
$3_c 2_L 1_R 1_{B-L}$ & monopoles & strings &  \begin{tabular} {l}
proton lifetime   ($Z_2$ broken)
\end{tabular}\\ \hline
\end{tabular}
\vspace{.5cm}

Table 1 : This is a table showing the formation of topological defects
in the possible symmetry breaking patterns from supersymmetric SO(10)
down to the standard model with broken matter parity. These models are
inconsistent with proton lifetime measurements. The table also
shows the relevant cosmological problems associated with each symmetry
breaking pattern, when occuring within a hybrid inflationary scenario.
 From a topological defect point of view, models with intermediate
$SU(5) \times \widetilde{U(1)}$, $3_c 2_L 2_R 1_{B-L}$ and
$3_c 2_L 1_R 1_{B-L}$ symmetry groups are compatible with
observations. The model with an intermediate $SU(5) \times
\widetilde{U(1)}$ symmetry is only possible in supergrativity
SO(10) models.

\vspace{1cm}
\begin{tabular}{|l|l|l|l|}
\hline
G & $SO(10) \rightarrow G$ & $G \rightarrow 3_c 2_L 1_Y Z_2$ & cosmological
problems \\ \hline
$SU(5) \times U(1)_X$ & monopoles-1 &  monopoles + $Z_2$-strings &
\begin{tabular} {l}
 monopoles-2
\end{tabular}\\ \hline
$SU(5) \times Z_2$ & $Z_2$-strings & monopoles-2 &  monopoles-2
\\ \hline
$SU(5) \times \widetilde{U(1)}$ & monopoles & $Z_2$-strings &
\begin{tabular} {l}
no problem, \\
monopoles inflated away
\end{tabular} \\ \hline
$4_c 2_L 2_R$ &  monopoles-1 & monopoles-2 + $Z_2$-strings &   monopoles-2\\
\hline
$4_c 2_L 2_R Z^c_2$ &\begin{tabular} {l}
  monopoles-1 \\
+ $Z_2$-strings
\end{tabular}
 &  \begin{tabular} {l}
 monopoles-2 + $Z_2$-strings\\
 + $Z_2$-walls
\end{tabular} &  \begin{tabular} {l}
 monopoles-2 +  $Z_2$-walls \end{tabular} \\ \hline
$3_c 2_L 2_R 1_{B-L}$ & monopoles & \begin{tabular} {l}
embedded strings \\
+ $Z_2$-strings
\end{tabular}&
\begin{tabular} {l}
no problem, \\
 monopoles inflated away
\end{tabular} \\ \hline
$3_c 2_L 2_R 1_{B-L} Z^c_2$ &  \begin{tabular} {l}
 monopoles \\
+ $Z_2$-strings  \end{tabular}
& \begin{tabular} {l}  embedded strings  + $Z_2$-strings \\
+ $Z_2$-walls
\end{tabular}& \begin{tabular} {l}   $Z_2$-walls \end{tabular} \\ \hline
$3_c 2_L 1_R 1_{B-L} $ & monopoles & $Z_2$-strings & \begin{tabular} {l}
no problem, \\
 monopoles inflated away
\end{tabular} \\ \hline
\end{tabular}
\vspace{.5cm}

Table 2 : This is a table showing the formation of topological defects
in the possible symmetry breaking patterns from supersymmetric SO(10)
down to the
standard model with unbroken matter parity. These models are
consistent with proton life time measurements and can
provide a superheavy Majorana mass to the right-handed neutrinos.
The table also shows the relevant cosmological problems associated
with each symmetry breaking pattern, when occurring within a
hybrid inflationary scenario. The
models with intermediate $SU(5) \times \widetilde{U(1)}$,
$3_c 2_L 2_R 1_{B-L}$ and $3_c 2_L 1_R
1_{B-L}$ symmetry groups are consistent with observations. The model with
intermediate $SU(5) \times \widetilde{U(1)}$ symmetry is only possible
in supergrativity SO(10) models.

\vspace{1cm}

\begin{tabular}{|c|c|}
\hline
$SO(10) \rightarrow 3_c 2_L 1_Y$ & cosmological problems\\ \hline
 monopoles-2 & \begin{tabular} {l}
monopoles-2 \\
+ proton lifetime ($Z_2$ broken)
\end{tabular} \\ \hline
\end{tabular}
\vspace{.5cm}

Table 3 : This is a table showing the formation of topological defects
in models where supersymmetric SO(10) breaks directly down to the
MSSM with broken matter parity. The table also shows the relevant
cosmological problems associated with the symmetry breaking pattern,
when occurring within a hybrid inflationary scenario. These models
are inconsistent with observations.

\vspace{1cm}
\begin{tabular}{|c|c|}
\hline
$SO(10) \rightarrow 3_c 2_L 1_Y Z_2$& cosmological problems \\ \hline
 monopoles-2 + $Z_2$-strings & monopoles-2 \\ \hline
\end{tabular}
\vspace{.5cm}

Table 4 : This is a table showing the formation of topological defects
in models where supersymmetric SO(10) breaks directly down to the MSSM
with unbroken matter parity. The table also shows the relevant
cosmological problems associated with the symmetry breaking pattern,
when occurring within a hybrid inflationary scenario. These models
are inconsistent with observations.


\vspace{1cm}



\begin{thebibliography}{99}
\bibitem[*]{byline} Also at King's College, Cambridge University.
\bibitem{Georgi} H. Georgi, Particles and Fields. Proceedings of the
APS Div. of Particles and Fields, ed. C. Carlson; H. Fritsch and
P. Minkowski, Ann. Phys. {\bf 93}, 193 (1975).
\bibitem{merge} S. Dimopoulos and H. Georgi, Nucl.\ Phys.\ {\bf B
193}, 150 (1981); J. Ellis, S. Kelley and D.V. Nanopoulos, Phys.\
Lett.\ {\bf B 249}, 441 (1990); U. Amaldi, W. de Boer and H.
Furstenau, Phys.\ Lett.\ {\bf B 260}, 447 (1991); P. Langacker and
M.X. Luo, Phys.\ Rev.\ {\bf D 44}, 817 (1991); M. Carena, S. Pokorski,
M. Olechowski and C.E.M. Wagner,
Nucl.\ Phys.\ B {\bf 406}, 59 (1993)
\bibitem{dimo} G. Anderson, S. Dimopoulos, L. Hall, S. Raby and
G. Starkman, Phys.\ Rev.\  {\bf D 49}, 3660 (1994); S. Raby, talk given
at SUSY 94, Warsaw, Poland; M. Carena, S. Dimopoulos, C.E.M. Wagner,
 S. Raby preprint CERN-TH-95-53; K.S. Babu and Q. Shafi, preprint
BA-95-05.
\bibitem{tanB} M. Olechowski and  S. Pokorski, Phys.\ Lett.\  {\bf B
214}, 393 (1988).
\bibitem{DW} S. Dimopoulos and F. Wilczek, Proceedings Erice Summer School, Ed.
A. Zichich (1981); Dae-Gya Lee and
R.N.Mohapatra, Phys.\ Lett.\ {\bf B 324}, 376, (1994).
\bibitem{seesaw} M. Gell-Mann, P. Ramond, and R. Slansky,
In Supergravity, edited by D. Freedman {\it et al.}  (North-Holland,
 Amsterdam, 1980); T. Yanagida, Proceedings of the KEK workshop, 1979
 (unpublished); R. N. Mohapatra and G. Senjan\'{o}vic. Phys.\ Rev.\
 Lett.\ {\bf 44}, 912 (1980).
\bibitem{MSW} S.P. Mikheyev and A.Y. Smirnov, Yad.\ Fiz.\, {\bf 42},
1441 (1985); L. Wolfenstein, Phys.\ Rev.\ {\bf D 17}, 2369 (1978);
  ibid.\ {\bf D 18}, 958 (1979);  ibid.\ {\bf D 20}, 2634 (1980).
\bibitem{baryon} M. Fukugita and T. Yanagida, Phys. Lett. {\bf B
174}, 45 (1986); H. Murayama, H. Suzuki and T. Yanagida, Phys. Rev. Lett. {\bf
70}, 1912 (1993).
\bibitem{Kibble} T.W.B. Kibble, J. Phys. {\bf A 9}, 387 (1976).
\bibitem{moha} R.N. Mohapatra and M.K. Parida, Phys.\ Rev.\ {\bf D
47}, 264 (1993); Dae-Gyu Lee, R.N. Mohapatra, M.K. Parida, and
M. Rani, University of Maryland preprint UMD-PP-94-117 (1994).
\bibitem{Srivastava} P.P. Srivastava, Lett.\ Nuovo Cimento {\bf 12},
161 (1976).
\bibitem{Albert}  A.A. Albert, Trans.\ ann.\ Math.\ Soc.\ {\bf 64}, 552 (1948).
\bibitem{Santilli} R.M. Santilli, Hadronic Journal {\bf 1}, 223
(1978).
\bibitem{Linde} A.D. Linde, Phys.\ Lett.\ {\bf B 259}, 38 (1991);
Phys.\ Rev.\ {\bf D 49}, 748 (1994).
\bibitem{Cop} E.J. Copeland, A.R. Liddle, D.H. lyth, E.D. Stewart and D. Wands,
Phys. Rev. {\bf D 49}, 6410 (1994).
\bibitem{Shafipot} G. Dvali, Q. Shafi and R. Schaefer, Phys. Rev. Lett. {\bf
73}, 1886 (1994).
\bibitem{Vil} A. Vilenkin, Nucl.\ Phys.\ {\bf B 196}, 240 (1982);
G. Lazarides, Q. Shafi and T. Walsh, Nucl.\ Phys.\ {\bf B 195}, 157
(1982).
\bibitem{Kib} T.W.B. Kibble, G. Lazarides and Q. Shafi, Phys.\ Rev.\
{\bf D 26}, 435, (1982); A. Vilenkin and A.E. Everett, Phys.\ Rev.\
Lett.\ {\bf 48}, 1867 (1982); A.E. Everett and A. Vilenkin, Nucl.\
Phys. {\bf B 207}, 43 (1982).
\bibitem{nano} J.P. Derendinger, J.E. Kim aand D.V. Nanopoulos, Phys. Lett.
{\bf 139B}, 170 (1984).
\bibitem{54} G. Lazarides, M. Magg and Q. Shafi, Phys. Lett. {\bf B 97}, 87
(1980); F. Bucella, L. Cocco and C. Wetterich, Nucl. phys. {\bf B 248}, 273
(1984).
\bibitem{210} D. Chang and A. Kumar, Phys. Rev. {\bf D 33}, 2695 (1986);
J. Basaq, S. Meljanac and L. O'Raifeartaigh, Phys. Rev. {\bf D 39}, 3310
(1989); X.-G. He and S. Meljanac, Phys. Rev. {\bf D 40}, 2098 (1989).
\bibitem{45} O. Kaymakcalan, L. Michel, K.C. Wali, W.D. McGlinn and
L. O'Raifeartaigh, Nucl. Phys. {\bf B 267}, 203 (1986); R. Thornburg
and W.D. McGlinn, Phys. rev. {\bf D 33}, 2991 (1986); R. Kuchimanchi,
Phys. Rev. {\bf D 47}, 685 (1993);
\bibitem{next} R. Jeannerot, preprint DAMTP 95-35.
\bibitem{Tanmay} T. Vachaspati, Phys.\ Rev.\ Lett.\ {\bf 68}, 1977 (1992).
\bibitem{so10cs}  M. Aryal and A. Everett, Phys. Rev. {\bf D 35}, 3105 (1987);
Chung-Pei Ma, Phys. Rev. {\bf D 48}, 530 (1993); A.C. Davis and R. Jeannerot,
Phys. Rev. D. {\bf 52}, 1944 (1995).
\bibitem{Nathan} A.C. Davis and N. Lepora, preprint DAMTP 95-05,
Phys. Rev. D (in press).
\bibitem{Martin}  S.P. Martin, Phys.\ Rev.\ {\bf D 46}, 2769 (1992).
\end{thebibliography}
\end{document}